\documentclass{ws-ijmpe}
\usepackage{amssymb}

\newcommand{\be}{\begin{equation}}
\newcommand{\ee}{\end{equation}}
\newcommand{\bea}{\begin{eqnarray}}
\newcommand{\eea}{\end{eqnarray}}

\newcommand{\re}[1]{Eq.~(\ref{#1})}

\newcommand{\ds}{\displaystyle}

\newcommand{\hsp}{\hspace*{1pt}}

\begin{document}

\markboth{Barbara~Betz et al.}
{Hydrodynamic Flow and Mach Cones}

\catchline{}{}{}{}{}
%%%%%%%%%%%%%%%%%%%%%%%%%%%%%%%%%%%%%%%%%%%%%%%%%%%%%%%%%%%%%%%%%%%%

\title{Mach Cones and Hydrodynamic Flow:\\
Probing Big Bang Matter in the Laboratory}

\author{\footnotesize Barbara~Betz$^{1,2}$, 
Philip~Rau$^1$, Horst~St\"ocker$^{1,3}$}

\address{$^1$Institut f\"ur Theoretische Physik, Johann Wolfgang
  Goethe - Universit\"at,\\
  Max-von-Laue Str.~1, 60438 Frankfurt, Germany\\
  $^2$Helmholtz Graduate School, GSI, FIAS and Universit\"at Frankfurt\\
  $^3$FIAS~-~Frankfurt Institute for Advanced Studies,\\
  Max-von-Laue Str.~1, 60438 Frankfurt, Germany}

\maketitle

\begin{history}
\received{(received date)}
\revised{(revised date)}
%\accepted{(Day Month Year)}
%\comby{(xxxxxxxxxx)}
\end{history}

\begin{abstract} %should be less than 200 words.
A critical discussion of the present signals for the phase transition
to quark-gluon plasma (QGP) is given. Since hadronic rescattering
models predict much larger flow than observed from 1 to 50
A GeV laboratory bombarding energies, this observation is interpreted
as potential evidence for a first-order phase transition at high
baryon density. A detailed discussion of the collective flow as a
barometer for the equation of state (EoS) of hot dense matter at RHIC
follows. Here, hadronic rescattering models can explain $< 30 \%$ of
the observed elliptic flow $v_2$ for $p_T > 2$ GeV/c. This is
interpreted as an evidence for the production of superdense matter at
RHIC. The connection of $v_2$ to jet suppression is examined. A study of Mach
shocks generated by fast partonic jets propagating through the QGP is
given. The main goal is to take into account different types of
collective motion during the formation and evolution of this matter. A
significant deformation of Mach shocks in central Au+Au collisions at
RHIC and LHC energies as compared to the case of jet propagation in a
static medium is predicted. A new hydrodynamical study of jet energy
loss is presented.
\end{abstract}

%--------------------------------------------------------------------
\section{Observables for the QGP phase transition}

Lattice QCD calculations yield a phase
diagram\cite{Fodor04,Karsch04} (see Fig. \ref{phasedia}) which
shows a crossing, but no first-order phase transition to the 
quark-gluon plasma (QGP) for vanishing or small chemical potentials
$\mu_B$, i.e. for conditions accessible at central rapidities at full
RHIC energy. A first-order phase transition is expected to occur only at high
baryochemical potentials or densities, i.e. at the lower SPS and RHIC
energies and in the fragmentation region of RHIC, $y \approx
3-5$\cite{Anishetty80,Date85}. Here, the critical
baryochemical potential is predicted\cite{Fodor04,Karsch04} to be large,
$\mu_B^c \approx 400 \pm 50 \mbox{ MeV}$, and the critical temperature
to be $T_c \approx 150-160$ MeV. We expect a first-oder phase transition also
at finite strangeness\cite{Greiner:1987tg}. Predictions for the phase
diagram of strongly interacting matter for realistic non-vanishing net
strangeness are urgently needed to obtain a comprehensive picture of
the QCD phase structure in all relevant dimensions (isospin,
strangeness, non-equilibrium) of the EoS. Multi-strange degrees of
freedom are very promising probes for the properties of the dense and
hot matter\cite{Koch86}.

\subsection{Thermodynamics in the $T$ - $\mu_B$ plane}

\begin{figure}[t]
\centerline{\epsfig{file=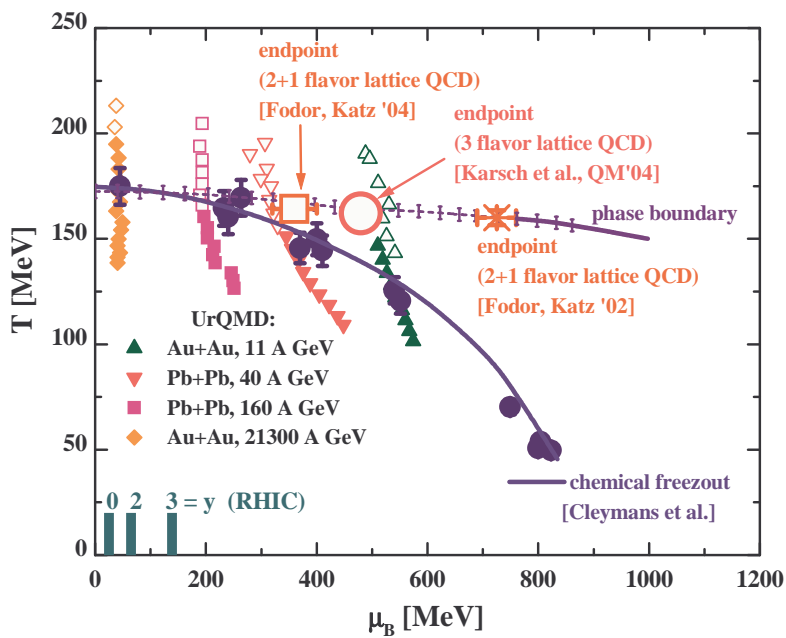,scale=0.55}}
\caption[]{The phase diagram with the critical end point at $\mu_B
  \approx 400 \mbox{ MeV}, T \approx 160 \mbox{ MeV} $ as predicted by
  Lattice QCD. In addition, the time evolution in the $T-\mu_B$--plane of
  a central cell in UrQMD calculations [from Bravina {\it et
    al.}]\cite{Bravina} is depicted for different bombarding
  energies. Note that the calculations indicate that bombarding
  energies $E_{Lab} \lesssim 40$ A GeV are needed to
  probe a first-order phase transition. At RHIC this point is
  accessible in the fragmentation region only [from Bratkovskaya {\it
    et al.}]\protect{\cite{Bratkov04}}.}
\label{phasedia}
\end{figure}

Figure \ref{phasedia} shows a comparison of the QCD predictions with
the thermodynamic parameters $T$ and $\mu_B$ extracted
from the UrQMD transport model in the central overlap regime of Au+Au
collisions\cite{Bratkov04}. Full dots with errorbars denote the 
'experimental' chemical freeze-out parameters -- determined from fits 
to the experimental yields -- taken from Ref.\cite{Cleymans}. Triangular and
quadratic symbols (time-ordered in vertical sequence) stand for
temperatures $T$ and chemical potentials $\mu_B$ extracted from UrQMD
transport calculations in central Au+Au (Pb+Pb) collisions at
RHIC\cite{Bravina} as a function of the reaction time (separated by 1
fm/c steps from top to bottom). Open symbols denote
nonequilibrium configurations and correspond to $T$ parameters
extracted from the transverse momentum distributions, whereas the full
symbols denote configurations in approximate pressure equilibrium in
longitudinal and transverse direction.

During the nonequilibrium phase (open symbols) the transport
calculations show much higher temperatures (or energy densities) than
the 'experimental' chemical freeze-out configurations at all bombarding
energies ($\geq$ 11 A GeV). These numbers are also higher than
the critical point (circle) of (2+1) flavor lattice QCD calculations
by the Bielefeld-Swansea-collaboration\cite{Karsch04} (large open
circle) and by the Wuppertal-Budapest-collaboration\cite{Fodor04} (the
star shows earlier results from\cite{Fodor04}). The energy density at
$\mu_c, T_c$ is of the order of $\approx$ 1 GeV/fm$^3$. 
At RHIC energies a cross-over is expected at midrapidity, when
the temperature drops during the expansion phase of the 'hot
fireball'. The baryon chemical potential $\mu_B$ has been obtained
from a statistical model analysis by the BRAHMS collaboration based on
measured antihadron to hadron ratios\cite{BRAHMS_PRL03} for different
rapidity intervals at RHIC energies. At midrapidity one finds
$\mu_B\simeq 0$, whereas at forward rapidities $\mu_B$ increases up to
$\mu_B\simeq 130$ MeV at $y=3$. Thus only a forward rapidity
measurement ($y \approx 4-5)$ will allow to probe large $\mu_B$ at
RHIC. The STAR and PHENIX detectors at RHIC offer 
a unique opportunity to reach higher chemical
potentials and the first-order phase transition region at
midrapidity in the HiMu-RHIC-running at $\sqrt{s}=4-12$ GeV in the coming year. 
The International FAIR Facility at GSI will be offering a fully devoted research program in the next
decade.

\subsection{Hydrodynamic flow}

Hydrodynamic flow and shock formation has been proposed
early\cite{Hofmann74,Hofmann76} as the key mechanism for the creation
of hot and dense matter in relativistic heavy-ion
collisions\cite{Lacey}. The full three-dimensional hydrodynamical flow
problem is much more complicated than the one-dimensional Landau
model\cite{Landau}: the 3-dimensional compression and expansion
dynamics yields complex triple differential cross-sections which
provide quite accurate spectroscopic handles on the EoS. The
bounce-off, $v_1(p_T)$ (i.e., the strength of the directed flow in the
reaction plane), the squeeze-out, $v_2(p_T)$ (the strength of the
second moment of the azimuthal particle emission
distribution)\cite{Hofmann74,Hofmann76,Stocker79,Stocker80,Stocker81,Stocker82,Stocker86},
and the
antiflow\cite{Stocker79,Stocker80,Stocker81,Stocker82,Stocker86}
(third flow component\cite{Csernai99,Csernai04}) serve as differential
barometers for the properties of compressed, dense matter from SIS to
RHIC. In particular, it has been
shown\cite{Hofmann76,Stocker79,Stocker80,Stocker81,Stocker82,Stocker86}
that the disappearance or ''collapse'' of flow is a direct result of a
first-order phase transition.

Several hydrodynamic models\cite{Rischke:1995pe} have been used in the
past, starting with the one-fluid ideal hydrodynamic approach. It is
well known that the latter model predicts far too large flow
effects. To obtain a better description of the dynamics, viscous fluid
models have been developed\cite{Schmidt93,Muronga01,Muronga03,Romatschke,Mota,HeinzChaudhuri,Hirano}. In
parallel, so-called three-fluid models, which distinguish between
projectile, target and the fireball fluid, have been
considered\cite{Brachmann97}. Here viscosity effects appear only
between the different fluids, but not inside the individual
fluids. The aim is to have at our disposal a reliable,
three-dimensional, relativistic three-fluid model including
viscosity\cite{Muronga01,Muronga03}.

Flow can be described very elegantly in hydrodynamics. However, also
consider microscopic multicomponent (pre-)hadron transport theory,
e.g. models like qMD\cite{Hofmann99}, IQMD\cite{Hartnack89}, UrQMD
\cite{Bass98}, or HSD\cite{Cassing99}, as control models for viscous
hydrodynamics and as background models to subtract interesting
non-hadronic effects from data. If hydrodynamics with and without
quark matter EoS, hadronic transport models without quark matter --
but with strings -- are compared to data, can we learn whether quark
matter has been formed? What degree of equilibration has been
reached? What does the EoS look like? How are the
particle properties, self-energies, cross sections changed?

\subsection{Review of AGS and SPS results}

\begin{figure}[t]
\epsfig{file=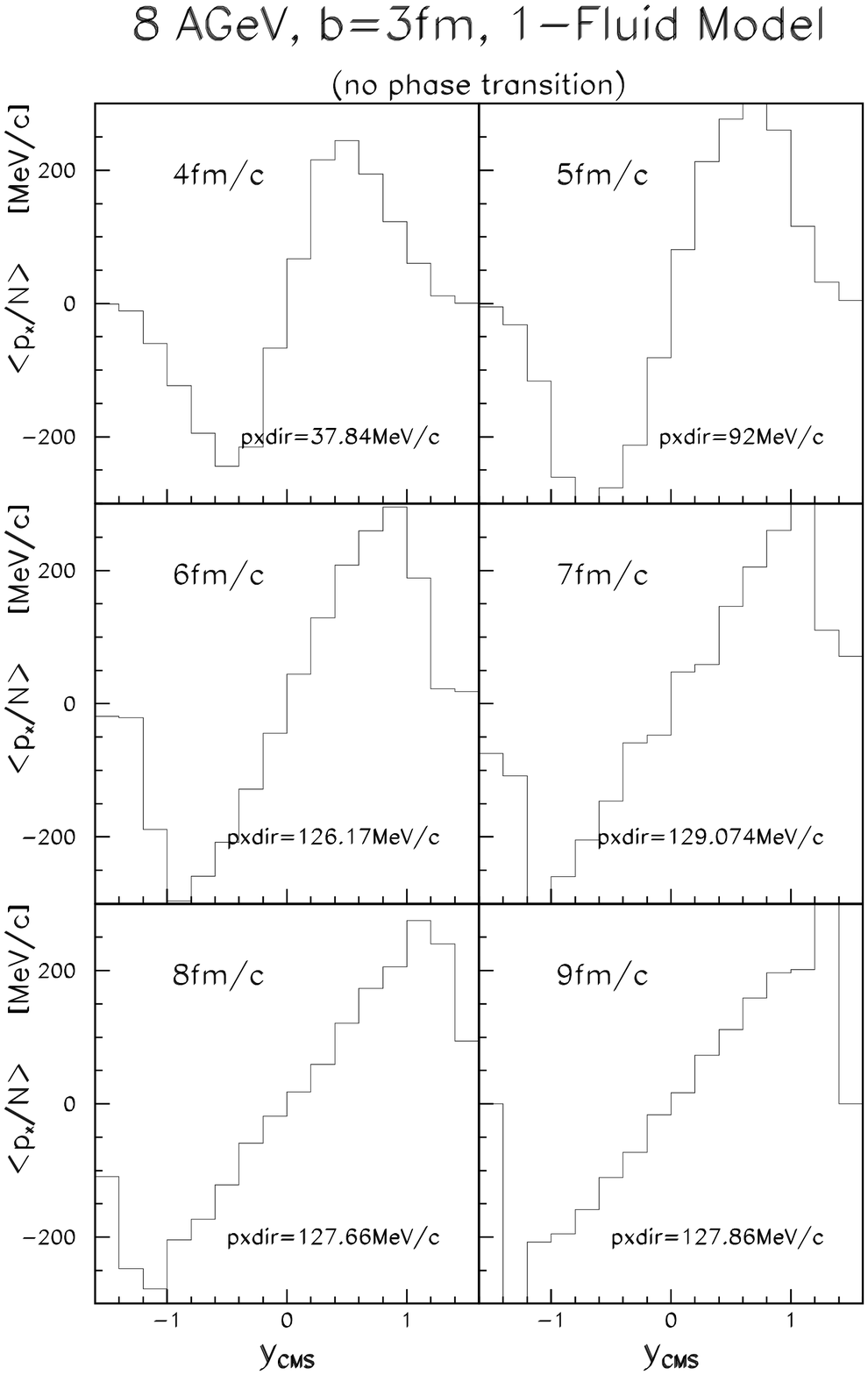,width=6.2cm}\hspace*{1.5mm}
\epsfig{file=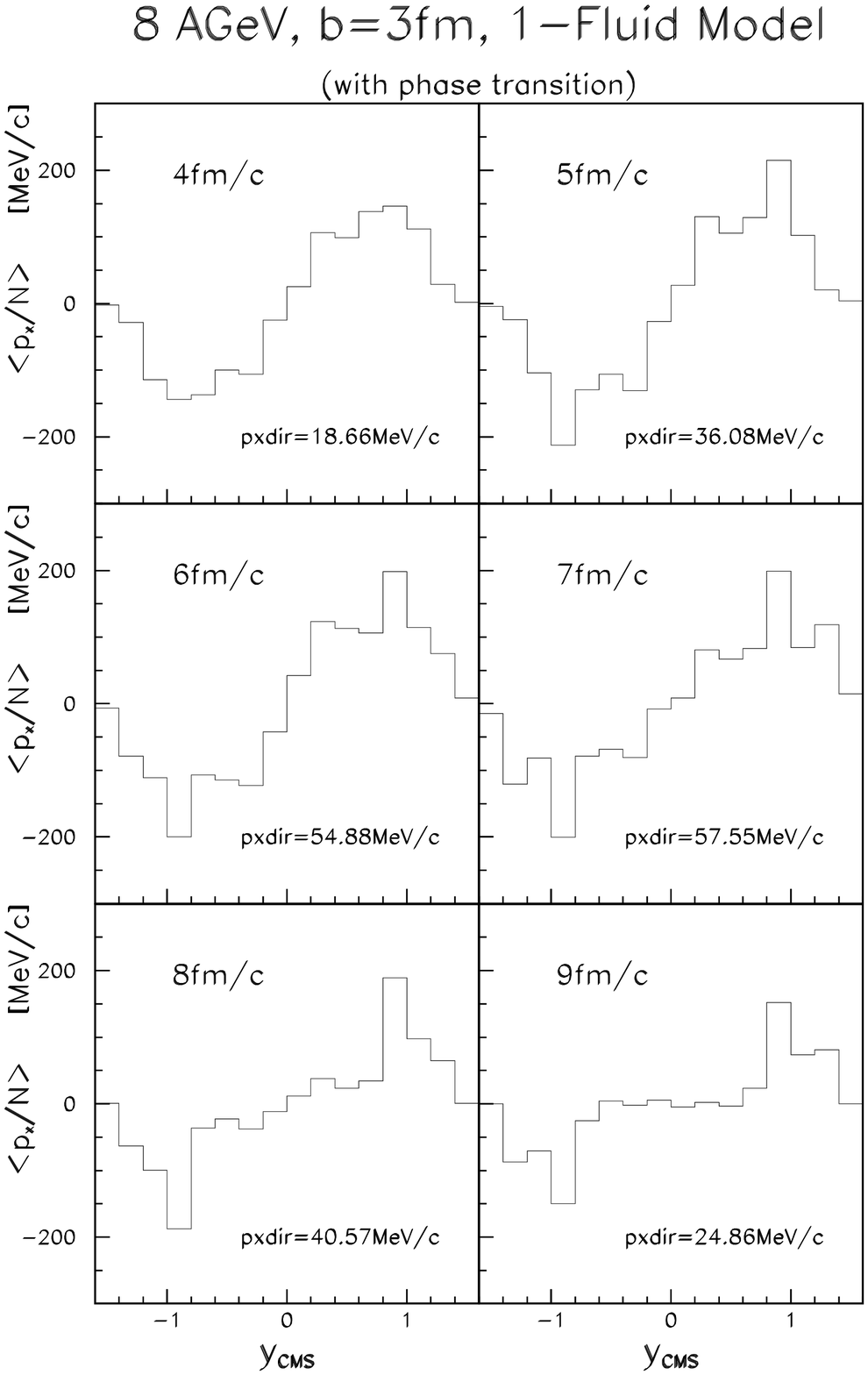,width=6.2cm}
\caption[]{Time evolution of directed flow $p_x/N$ as a function of
  rapidity for Au+Au collisions at 8 A GeV in the one-fluid
  model. Left: Hadronic EoS without phase transition. Right: An EoS
  including a first-order phase transition to the QGP [from
  Brachmann]\protect{\cite{Brach00}}.
\label{flow_brach1}} 
\end{figure}

Microscopic (pre-)hadronic transport models describe the formation and
distributions of many hadronic particles at AGS and SPS rather
well\cite{Weber02}. Furthermore, the nuclear EoS has
been extracted by comparing to flow data which are described
reasonably well up to AGS
energies\cite{Csernai99,Andronic03,Andronic01,Soff99,Sahu1,Sahu2}.
Ideal hydrodynamical calculations, on the other hand, predict far too
much flow at these energies\cite{Schmidt93}. Thus, viscosity effects
have to be taken into account.

\noindent
In particular, ideal hydrodynamical calculations yield factors of two
higher for the sideward flow at SIS\cite{Schmidt93} and AGS, while the
directed flow $p_x/m$ measurement of the E895 collaboration shows that
the $p$ and $\Lambda$ data are reproduced reasonably well\cite{Soff99,Sto04}
in UrQMD, i.e., in a hadronic transport
theory with reasonable cross-sections, i.e. realistic mean-free-path
of the constituents.

\begin{figure}[t]
\centerline{\epsfig{file=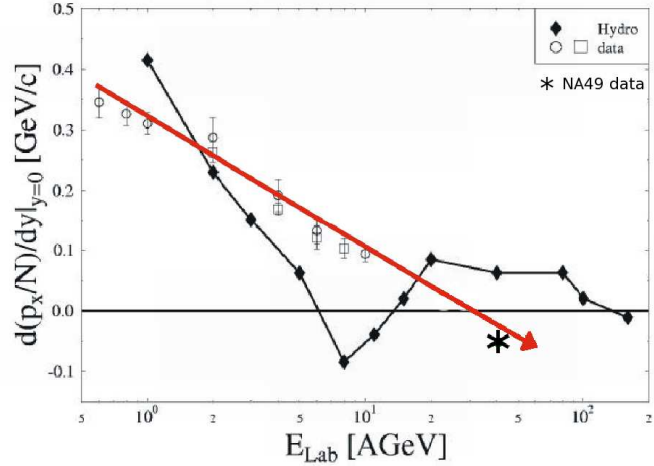,scale=0.38}}
\caption[]{Measured SIS and AGS proton $dp_x/dy$-slope data compared
  to a one-fluid hydrodynamical calculation. A linear extrapolation of
  the AGS data indicates a collapse of flow at $E_{Lab} \approx 30$ A
  GeV [see also Ref.\protect{\cite{Brach99}}], i.e. for the lowest SPS- and 
  the upper FAIR- energies at GSI
  [from Paech {\it et al.}]\protect{\cite{Paech00}. The point at $40$~A GeV
    is calculated using the NA49 central data, cf. Fig. \ref{sps_v1_data}.}
\label{flow_extra}}
\end{figure}

Only ideal hydrodynamical calculations predict, however, the appearance of a
so-called ''third flow component''\cite{Csernai99} or
''antiflow''\cite{Sto04,Brach00} in central collisions. We stress that
this only holds if the matter undergoes a first-order phase transition
to the QGP. The signal is that around midrapidity the directed flow,
$p_x (y)$, of protons develops a negative slope! In contrast, a
hadronic EoS without QGP phase transition does not yield such an
exotic ''antiflow'' (negative slope) wiggle in the proton flow
$v_1(y)$. The ideal hydrodynamic time evolution of the directed flow,
$p_x/N$, for the purely hadronic EoS (Fig. \ref{flow_brach1} l.h.s.)
does show a clean linear increase of $p_x(y)$, just as the microscopic
transport theory and as the data\cite{Soff99}. For an EoS including a
first-order phase transition to the QGP (Fig. \ref{flow_brach1}
r.h.s.) it can be seen, however, that the proton flow $v_1 \sim
p_x/p_T$ collapses; the collapse occurs around midrapidity. This
observation is explained by an antiflow component of protons, which
develops when the expansion from the plasma sets in\cite{Brach99}.

The ideal hydrodynamic directed proton flow $p_x$ (Fig.
\ref{flow_extra}) shows even negative values between 8 and 20
A GeV. An increase back to positive flow is predicted with
increasing energy, when the compressed QGP phase is probed. But, where
is the predicted minimum of the proton flow in the data? The
hydrodynamical calculations suggest this ''softest-point collapse'' is
at $E_{Lab} \approx 8$ A GeV. This has not been verified by the AGS
data! However, a linear extrapolation of the AGS data indicates a
collapse of the directed proton flow at $E_{Lab} \approx 30$ A GeV
(Fig. \ref{flow_extra}).

\begin{figure}[t]
\begin{center}
\epsfig{file=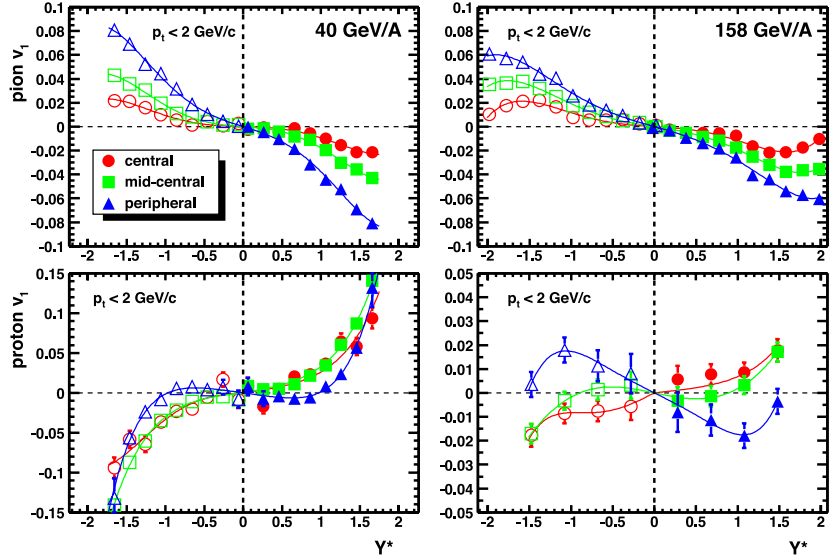,scale=0.62}
\caption[]{$v_1$ at SPS, 40 A GeV and 158 A GeV [from Alt {\it et
    al.}]\protect{\cite{NA49_v2pr40}}. The proton antiflow is observed
  in the NA49 experiment even at near central collisions, which is in
  contrast to the UrQMD model involving no phase transition
  (Fig.\protect\ref{v1_sps40}).} 
\label{sps_v1_data}
\end{center}
\end{figure}

\begin{figure}[t]
\begin{center}
\centerline{\epsfig{file=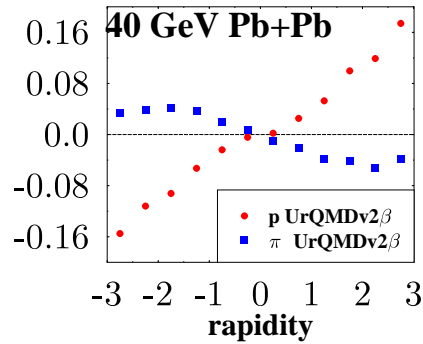,scale=0.34}}
\caption[]{Proton and pion flow $v_1=p_x/p_T$ at 40 A GeV as
  obtained within the UrQMD model. No proton antiflow is generated in
  this hadronic transport theory without phase transition (c.f.\ Ref.\ \protect{\cite{Petersen}}).}
\label{v1_sps40}
\end{center}
\end{figure}

Recently, substantial support for this prediction has been obtained by
the low energy 40 A GeV SPS data of the NA49
collaboration\cite{NA49_v2pr40,Petersen} (cf. Fig. \ref{sps_v1_data}). These
data clearly show the first proton ''antiflow'' around mid-rapidity,
in contrast to the AGS data as well as to the UrQMD calculations
involving no phase transition (Fig. \ref{v1_sps40}). Thus, at
bombarding energies of 30-40 A GeV, a first-order phase transition to
the baryon-rich QGP is most likely observed; hence the first order phase
transition line is crossed (cf. Fig. \ref{phasedia}). This is the
energy region where the new FAIR facility at GSI will operate. There
are good prospects that the baryon flow collapses and other
first-order QGP phase transition signals can be studied soon at the
lowest SPS energies as well as at the RHIC 
planned HiMu-run at midrapidity as well as the fragmentation region $y >
4-5$ for the highest RHIC and LHC-collider energies. 
These experiments will enable a detailed study of the
first-order phase transition at high $\mu_B$ and of the properties of
the baryon-rich QGP in the near future.

%--------------------------------------------------------------------
\section{Proton elliptic flow collapse at 40 A GeV - evidence for a
first-order phase transition at highest net baryon densities}

%pr  Bild rausgenommen
%
% \begin{figure}[h]
% \centerline{\epsfig{file=v2y-pb40.eps,scale=0.35}}
% \caption[]{Elliptic flow $v_2$ of protons (lower frame) and pions
% (upper frame) versus rapidity for Pb+Pb collisions at 40 A GeV
% from UrQMD calculations [from Soff {\it et al.}]\protect{\cite{Soff99}}.}
% \label{soff_v2pp40}
% \end{figure}

\begin{figure}[t]
\centerline{\epsfig{file=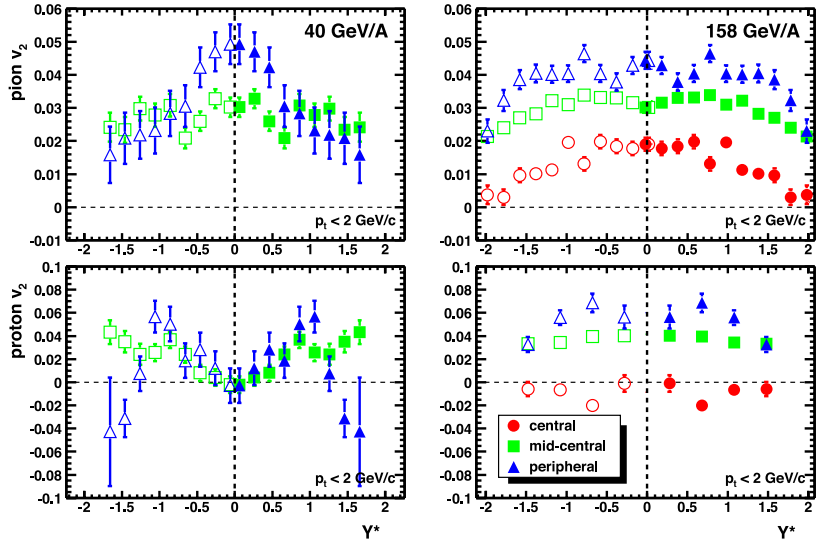,scale=0.6}}
\caption[]{Elliptic flow $v_2$ of protons versus rapidity at 40
  A GeV Pb+Pb collisions [from Alt {\it et
    al.}]\protect\cite{NA49_v2pr40} as measured by NA49 for three
  centrality bins: central (dots), mid-central (squares) and
  peripheral (triangles).}
\label{Fig_v2pr40}
\end{figure}

At SIS energies, microscopic transport models reproduce the data on
the excitation function of the proton elliptic flow $v_2$ quite well:
A soft, momentum-dependent EoS\cite{Andronic00,Andronic99} seems to account for the
data. The observed proton flow $v_2$ below $\sim$ 5 A GeV is smaller
than zero, which corresponds to the squeeze-out predicted by
hydrodynamics long
ago\cite{Hofmann74,Hofmann76,Stocker79,Stocker80,Stocker81,Stocker82,Stocker86}.
The AGS data exhibit a transition from
squeeze-out to in-plane flow in the midrapidity region. The change in
sign of the proton $v_2$ at 4-5 A GeV is in accord with transport
calculations (UrQMD calculations\cite{Soff99} for HSD
results see Ref.\cite{Sahu1,Sahu2}). At higher energies, 10-160 A GeV,
a smooth increase of the flow $v_2$ is predicted from the hadronic
transport simulations. In fact, the 158 A GeV data of the NA49
collaboration suggest that this smooth increase proceeds between AGS
and SPS as predicted. Accordingly, UrQMD calculations without phase
transition give considerable 3\% $v_2$ flow for midcentral and
peripheral protons at 40 A GeV (Ref.\cite{Sto04,Soff99}).

This is in strong contrast to recent NA49 data at 40 A GeV\cite{NA49_v2pr40,Petersen} (cf.
Fig. \ref{Fig_v2pr40}): A sudden collapse of the proton flow is
observed for midcentral as well as for peripheral protons. This
collapse of $v_2$ for protons around midrapidity at 40 A GeV is
very pronounced while it is not observed at 158 A GeV.

A dramatic collapse of the flow $v_1$ is also observed by
NA49\cite{NA49_v2pr40,Petersen}, again around 40 A GeV, where the collapse of
$v_2$ has been observed. This is the
highest energy - according to Ref.\cite{Fodor04,Karsch04} and
Fig. \ref{phasedia} - at which a first-order phase transition can be
reached at the central rapidities of relativistic heavy-ion
collisions. We therefore conclude that a first-order phase transition
at the highest baryon densities accessible in nature has been seen at
these energies in Pb+Pb collisions. Moreover, Ref.\cite{Paech03} shows
that the elliptic flow clearly distinguishes between a first-order
phase transition and a crossover.

\section{ Mach shocks induced by partonic jets in expanding QGP}
\label{sec1}

Sideward peaks have been recently
observed\cite{StarAngCorr,Adl03b,Wan04,Jac05} in azimuthal
distributions of secondaries associated with the high-$p_T$ hadrons in
central Au+Au collisions at \mbox{$\sqrt{s}=200$\,GeV}. In
Ref.\cite{Sto04} such peaks had been predicted as a signature of Mach
shocks created by partonic jets propagating through a QGP formed in
heavy--ion collisions. Analogous Mach shock waves were studied
previously in cold hadronic
matter\mbox{\cite{Hofmann74,Stocker79,Stocker86,Rischke90,Cha86}} as
well as in nuclear Fermi liquids\cite{Gla59,Kho80}. Recently, Mach
shocks from jets have been studied in Ref.\cite{Cas04,Ruppert,Chaudhuri}.

It is well known\cite{Landau} that a point--like perturbation (a small
body, a hadron or parton etc.) moving with supersonic speed in the
spatially homogeneous ideal fluid produces the so--called Mach region
of the perturbed matter. In the fluid rest frame (FRF) the Mach region
has a conical shape with an opening angle with respect to the direction
of particle propagation given by the expression\footnote{Here and
  below quantities in the FRF are marked by tilde.} 
$
\widetilde{\theta}_M=\sin^{-1}\left(\frac{c_s}{\widetilde{v}}\right)\,,
$
where $c_s$ denotes the sound velocity of the unperturbed (upstream)
fluid and $\widetilde{\bm{v}}$ is the particle velocity with respect to the
fluid. In the FRF, trajectories of fluid elements (perpendicular to the
surface of the Mach cone) are inclined at the angle
$\Delta\theta=\pi/2-\widetilde{\theta}_M$ with respect to
$\widetilde{\bm{v}}$\,. Strictly speaking, the above formula is
applicable only for weak, sound--like perturbations. It is certainly
not valid for space--time regions close to a leading particle.
Nevertheless, we shall use this simple expression for a qualitative
analysis of flow effects\cite{Satarov}. Following Refs.\cite{Sto04,Cas04} one can
estimate the angle of preferential emission of secondaries associated with
a fast jet in the QGP. Assuming the particle velocity to be
$\widetilde{v}=1$ and the sound velocity to be $c_s=1/\sqrt{3}$ leads to
$\Delta\theta\simeq 0.96$\,. This agrees well with positions of maxima
of the away--side two--particle distributions observed in central
Au+Au collisions at RHIC energies.

\section{ Deformation of Mach shocks due to radial flow}
\label{sec2}

Assuming that the away--side jet propagates with
velocity $\bm{v}$ parallel to the matter flow velocity $\bm{u}$\, and
$\bm{u}$ does not change with space and time, one sees that after
performing the Lorentz boost to the FRF, a weak Mach
shock has a conical shape with the axis along $\bm{v}$\,. In this
reference frame, the shock front angle $\widetilde{\theta}_M$ is again given
by $\widetilde{\theta}_M=\sin^{-1}\left(\frac{c_s}{\widetilde{v}}\right)\,$.  
Transformation from the FRF to the center of mass frame (CMF)
shows that the Mach region remains conical, but the Mach angle becomes
smaller in the CMF, 
$\tan{\theta_M}=\frac{\ds 1}{\ds\gamma_u}\tan{\widetilde{\theta}_M}\,,$
where $\gamma_u\equiv (1-u^2)^{-1/2}$ is the Lorentz factor
corresponding to the flow velocity~$\textbf{u}$\,.
Using the above Eqns.\ leads to the expression
for the Mach angle in the CMF
\begin{equation}\label{macp2}
\theta_M=\tan^{-1}
\left(c_s\sqrt{\frac{1-u^2}{\widetilde{v}^{\hsp 2}-c_s^2}}\right)\,,
\end{equation}
where
\begin{equation}\label{vrel}
\widetilde{v}=\frac{v\mp u}{1\mp v\hsp u}\,,
\end{equation}
and the upper (lower) sign corresponds to the jet's motion in (or opposite to)
the direction of collective flow. For ultrarelativistic jets ($v\to 1$) one
can take $\widetilde{v}\simeq 1$ which leads to a simpler expression
\begin{equation}\label{macp3}
\theta_M\simeq\tan^{-1}\left(\frac{\ds c_s\gamma_s}{\ds \gamma_u}\right)=
\sin^{-1}\left(c_s\sqrt{\frac{1-u^2}{1-u^2\hsp c_s^2}}\right)\,,
\end{equation}
with $\gamma_s=(1-c_s^2)^{-1/2}$\,. According to \re{macp3},
in the ultrarelativistic limit $\theta_M$ does not depend on the 
direction of flow with respect to the jet. The Mach cone 
becomes more narrow as compared to jet propagation in static matter. This
narrowing effect has a purely relativistic origin. Indeed, the
difference between~$\theta_M$ from \re{macp3} and the Mach angle in
absence of flow ($\lim\limits_{u\to 0}{\theta_M}=\sin^{-1}{c_s}$) is of
second order in the collective velocity $u$\,.

The case of a jet propagating at nonzero angle with respect to the flow
velocity is more complicated. Mach shocks
become nonconical for non--collinear flows. For simplicity, we
study only the case when the jet and flow velocities are
orthogonal to each other, $\bm{v}\perp\bm{u}$. Let axes $OX$ and $OY$
be directed along $\bm{u}$ and $\bm{v}$\,, respectively. We
first make the transition to the FRF by performing a Lorentz boost along
the $OX$ axis which leads to a jet velocity $\widetilde{v}$. 

\begin{figure}[t]
\begin{center}
\vspace*{-6.4cm}
\epsfig{file=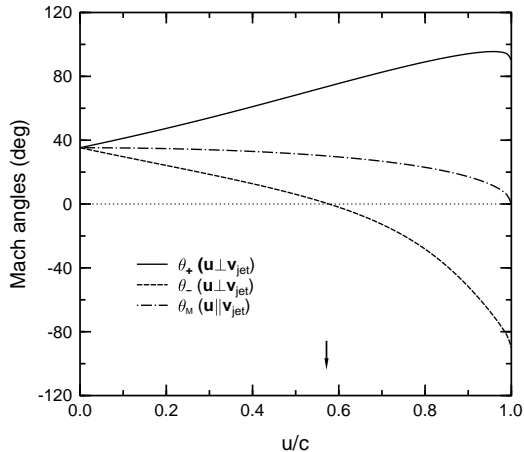,width=0.65\textwidth}
%\vspace*{-1cm}
\caption[]{Angles of Mach region created by a jet moving transversely
  (solid and dashed curves) and collinearly (dashed--dotted line) to
  the fluid velocity~$\bm{u}$\, in the CMF. All curves correspond to
  the case~\mbox{$c_s^2=1/3$}\,. The arrow marks the value $u=c_s$\,
  [from Satarov et al.]\protect{\cite{Satarov}}.}
\label{fig3}
\end{center}
\end{figure}

%%%%%%%%%%%%%%%%%%%%%%%%%%%%%%%%%%%%%%%%%%%%%%%%%%%%%%%%%%%%%%
%pr folgender Abschnitt nach Herausnahme von fig2 überabeitet:
%pr könnte unverständlich sein!
%  Figure~\ref{fig2} illustrates 
%%%%%%%%%%%%%%%%%%%%%%%%%%%%%%%%%%%%%%%%%%%%%%%%%%%%%%%%%%%%%
Assume a jet propagating along the
path $OA=\widetilde{v}\hsp\widetilde{t}$ during the time interval
$\widetilde{t}$ in the FRF. At
the same time, the wave front from a point--like perturbation (created
at point $O$) reaches a spherical surface with radius
$OB=OC=c_s\widetilde{t}$. Two tangent lines $AB$ and $AC$ show
the boundaries of the Mach region\footnote{Such region exists only if
  $\widetilde{v}>c_s$. This condition is fulfilled if $v>c_s$
  or $u>c_s$ hold.}
with the symmetry axis $OA$\,. This region has a conical shape with
opening angles $\widetilde{\theta}$ determined by the expressions
$\sin{\widetilde{\theta}}=\frac{OC}{OA}=\frac{\ds c_s}{\ds\widetilde{v}}
\simeq{c_s}\,.$

Performing inverse transformation from FRF to CMF, it is easy to show
that the Mach region is modified in two ways. First, it is no longer
symmetrical with respect to the jet trajectory in the CMF. 
%%%%%%%%%%%%%%%%%%%%%%%%%%%%%%%%%%%%%%%%%
%The insert in Fig.~\ref{fig2} shows that
%%%%%%%%%%%%%%%%%%%%%%%%%%%%%%%%%%%%%%%%% 
The boundaries of the Mach wave have different angles,
$\theta_+\neq\theta_-$, with respect to $\bm{v}$ in this reference
frame. One can interpret this effect as a consequence of transverse
flow which acts like a ''wind'' deforming the Mach cone along the
direction $OX$. On the other hand, the angles of the Mach front with
respect to the beam ($OZ$) axis are not changed under the
transformation to the CMF. We conclude that, due to effects of
transverse flow, the Mach region in the CMF should have a shape of a
deformed cone with an elliptic base. Figure~\ref{fig3} shows the
numerical values of the Mach angles for an ultrarelativistic jet
moving through the QGP transversely or collinearly to its flow velocity.

We point out a much stronger sensitivity of the
Mach angles $\theta_\pm$ to the transverse flow velocity
as compared with the collinear flow.

\begin{figure}[t]
\vspace*{-6.5cm}
\hspace*{3cm}\epsfig{file=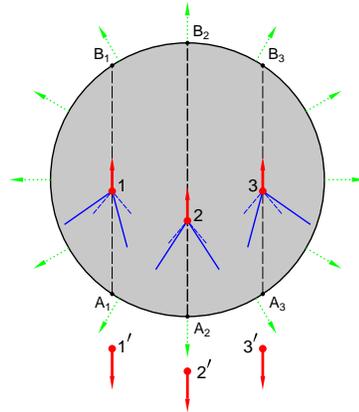,width=0.65\textwidth}
\vspace*{-1cm}
\caption[]{Schematic picture of Mach shocks from jets $1,2,3$
  propagating through the fireball matter (shaded circle) created in a
  central heavy--ion collision. Dotted arrows represent local
  velocities of the fireball expansion. Thick downward arrows show
  associated trigger jets. The Mach shock boundaries are shown by
  solid lines. Short--dashed lines give the positions of the shock fronts in
  the case of a static fireball [from Satarov et
  al.]\protect{\cite{Satarov}}.}
\label{fig4}
\end{figure}

To discuss possible observable effects, in Fig.~\ref{fig4}
we schematically show events with different di--jet
axes $A_iB_i\,(i=1,2,3)$\, with respect to the center of a 
fireball\footnote{ For simplicity we consider the case when both
  trigger ($i^\prime$) and away--side ($i$) jets have zero
  pseudorapidities in the CMF.}. 
In the $2-2^\prime$ event, the away--side jet '2' propagates along the
diameter $A_2B_2$\,, i.e. collinearly with respect to the collective
flow. In the two other cases, the di--jet axes are oriented along the chords,
$A_1B_1$ and $A_3B_3$\,, respectively. In such events, the fluid
velocity has both transverse and collinear components with respect to
the jet axis. We also show how the
Mach fronts will be deformed in an expanding matter. It is easy to see
that the radial expansion of the fireball should cause a broadening of
the sideward peaks in the $\Delta\phi$--distributions of associated
hadrons. Due to the radial expansion, the peaks will
acquire an additional width of the order of
\mbox{$<\theta_+-\theta_->$}\,. Here $\theta_\pm$ are local values of
the Mach angles in individual events. The angular brackets mean
averaging over the jet trajectory in a given event as well as over all
events with different positions of di--jet axes. Assuming that particle
emission is perpendicular to the surface of Mach cone and taking
$<u>\sim 0.4, c_s\simeq 1/\sqrt{3}$\,, we estimate the angular spread
of emitted hadrons in the range $30^\circ-50^\circ$\,. This is
comparable with the half distance between the away--side peaks of the
$\Delta\phi$ distribution observed by the STAR and PHENIX
collaborations\cite{StarAngCorr,Adl03b,Wan04,Jac05}. On the basis of this
analysis we conclude that in individual events the sideward maxima
should be asymmetric and more narrow than in an ensemble of
different events. Due to a stronger absorption of particles
emitted from the inner part of the shock (events 1, 3 in 
Fig.~\ref{fig4}), the two peaks may have different amplitudes.  
We think that these effects can be observed by
measuring three--particle correlations.

There is one more reason for broadening of the 
$\Delta\phi$--distributions which one should keep in mind when comparing
with experimental data: due to the momentum
spread of the initial parton distributions, $\Delta p_*\lesssim 1$\,GeV,
the di--jet system has a nonzero total momentum with respect to the
global CMF. As a consequence, the angle $\theta_*$
between the trigger-- and the away--side jet is generally speaking
not equal to $\pi$\,, as was assumed above. Taking typical momenta of
initial partons as $p_0$\,, with 
$p_0>4-6$\,GeV\cite{StarAngCorr,Adl03b,Wan04,Jac05}, we estimate the angular 
spread as $|\pi-\theta_*|\sim\Delta p_*/p_0\lesssim 0.1$\,. 
Therefore, the considered broadening should be much less than the typical
shift of the Mach angles due to the collective flow.

%--------------------------------------------------------------------
\section{Angular Correlations of Jets -- Can jets fake the large
$v_2$-values observed?}\label{sectionjets}

\begin{figure}[t]
\centerline{\psfig{file=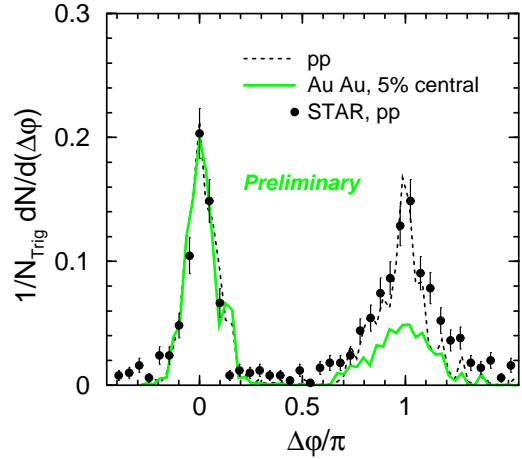,angle=-90,width=6.7cm}}
\caption[]{STAR data on near-side and away--side jet correlation
  compared to the HSD model for p+p and central Au+Au collisions at
  midrapidity for $p_T(N_{Trig})=4\dots6\,{\rm GeV}/c$ and
  $p_T=2\,{\rm GeV}/c\dots p_T(N_{Trig})$ [from Cassing {\it et
    al.}]\protect{\cite{Gal04,CGG}}.}
    \label{angcorr}
\end{figure}

\begin{figure}[t]
\centerline{\psfig{file=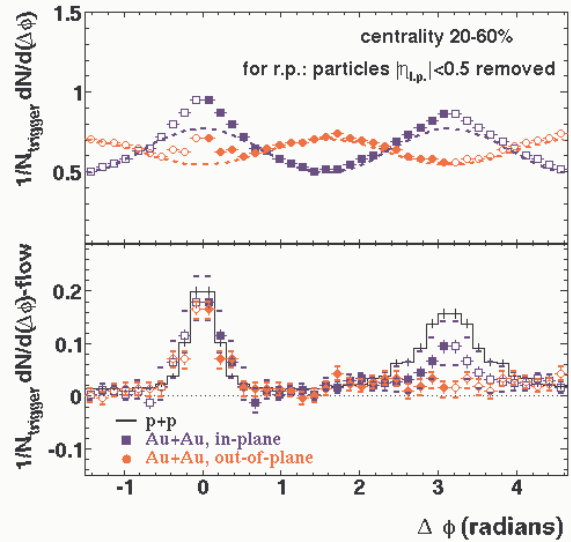,width=7.9cm}}
\caption[]{High $p_T$ correlations: in-plane vs. out-of-plane
  correlations of the probe (jet+secondary jet fragments) with the
  bulk ($v_2$ of the plasma at $p_T > 2\,$GeV/c) prove the existence
  of the initial plasma state (STAR-collaboration, preliminary).}
\label{filimonov}
\end{figure}

Figure \ref{angcorr} shows the angular correlation of high-$p_T$
particles ($p_T \rm{(N_{Trig})}=4\dots6\,{\rm GeV}/c$, $p_T=2\,{\rm
  GeV}\dots p_T \rm{(N_{Trig})}$, $|y| <0.7$) for the 5\% most central
Au+Au collisions at $\sqrt{s}$ = 200 GeV (solid line) as well as $p+p$
reactions (dashed line) from the HSD-model\cite {Gal04} in comparison
to the data from STAR for $p+p$ collisions\cite{StarAngCorr}. Gating on
high-$p_T$ hadrons (in the vacuum) yields 'near--side' correlations in
Au+Au collisions close to the 'near--side' correlations observed for
jet fragmentation in the vacuum (p+p). This is in agreement with the
experimental observation\cite{StarAngCorr,Phenix}. However, for the away--side
jet correlations, the authors of Ref.\cite {Gal04} get only a
$\sim$50\% reduction, similar to HIJING, which has only parton
quenching and neglects hadron rescattering. Clearly, the
observed\cite{StarAngCorr} complete disappearance of the away--side jet
(see Fig.~\ref{filimonov}) cannot be explained in the HSD-(pre-)hadronic
cascade even with a small formation time of $0.8\,$fm/c. Hence, the
correlation data provide another clear proof for the existence of the
bulk plasma.

Although (pre-)hadronic final-state interactions yield a sizable ($\leq
50 \%$) contribution to the high-$p_T$ suppression effects observed in
Au+Au collisions at RHIC, $\sim 50 \%$ of the jet suppression
originates from interactions in the plasma phase. The elliptic flow,
$v_2$, for high-transverse momentum particles is underestimated by at
least a factor of 3 in the HSD transport calculations\cite{CGG}. 
The experimentally observed proton excess over
pions at transverse momenta $p_T > 2.5$ GeV/c cannot be explained
within the CGG approach\cite{CGG}; in fact, the proton yield at
high-$p_T \geq 5$ GeV/c is a factor 5-10 too small. We point out that
this also holds for partonic jet-quenching models. Further
experimental data on the suppression of high-momentum hadrons from
d+Au and Au+Au collisions, down to $\sqrt{s}$ = 20 GeV, are
desperately needed to separate initial-state Cronin effects from
final-state attenuation and to disentangle the role of partons in the
colored parton plasma from those of interacting pre-hadrons in the hot
and dense fireball.

Can the attenuation of jets of $p_T \ge5\,$GeV/c actually fake the
observed $v_2$-values at $p_T \approx 2\,$GeV/c? This question comes
about since due to fragmentation and rescattering a lot of
momentum-degraded hadrons will propagate in the hemisphere defined by
the jets. However, their momentum dispersion perpendicular to the jet
direction is so large that it could indeed fake a collective flow that
is interpreted as coming from the early high-pressure plasma phase.

\begin{figure}[t]
\centerline{\psfig{file=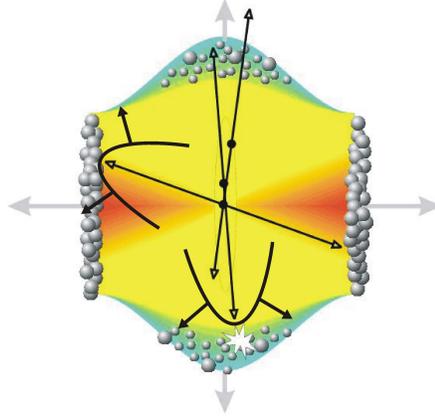,width=5.8cm}}
\caption[]{Illustration of jets traveling through the late hadronic
  stage of the reaction. Only jets from the region close to the
  initial surface can propagate and fragment in the
  vacuum\protect{\cite{Hofmann74,Gal04,Baum75}}. The other
  jets will interact with the bulk, resulting in wakes with bow waves
  travelling transversely to the jet axis.}
\label{fig:scheme}
\end{figure}

On first sight, Fig. \ref{filimonov} shows that this could indeed be
the case: the in-plane $v_2$ correlations are aligned with the jet
axis, the away--side bump, usually attributed to collective $v_2$ flow
(dashed line), could well be rather due to the stopped, fragmented and
rescattered away--side jet! However, this argument is falsified by the
out-of-plane correlations (circles in Fig. \ref{filimonov}). The
near-side jet is clearly visible in the valley of the collective flow
$v_2$ distribution. Note that $v_2$ peaks at $\varphi = \pi/2$
relative to the jet axis! The away--side jet, on the other hand, has
completely vanished in the out-of-plane distribution
(cf. Fig. \ref{fig:scheme}).

Where are all the jet fragments gone and why is there no trace left?
Even if the away--side jet fragments completely and the fragments get
stuck in the plasma, leftovers should be detected at momenta below
$2\,$GeV/c. Hadronic models as well as parton cascades will have a
hard time to get a quantitative agreement with these exciting data.

We propose future correlation measurements which can yield
spectroscopic information on the plasma:

\begin{enumerate}
\item
If the plasma is a color-electric plasma\cite{Sto04,Ruppert:2005uz}, 
experiments will - in spite of
strong plasma damping - be able to search for wake-riding potential
effects. The wake of the leading jet particle can trap comoving
companions that move through the plasma in the wake pocket with the
same speed ($p_T/m$) as the leading particle. This can be particular
stable for charmed jets due to the deadcone effect as proposed by
Kharzeev et al\cite{Kharzeev}, which will guarantee little energy
loss, i.e. constant velocity of the leading D-meson. The leading
D-meson will practically have very little momentum degradation in the
plasma and therefore the wake potential following the D will be able to
capture the equal speed companion, which can be detected\cite{Schafer78}.

\item
One may measure the sound velocity of the expanding plasma by the
emission pattern of the plasma particles travelling sideways with
respect to the jet axis: The dispersive wave generated by the wake of
the jet in the plasma yields preferential emission to an angle
(relative to the jet axis) which is given by the ratio of the leading
jet particles' velocity, devided by the sound velocity in the hot dense
plasma rest frame. The speed of sound for a non-interacting gas of
relativistic massless plasma particles is $c_s \approx
\frac{1}{\sqrt{3}} \approx 57 \% \,c$, while for a plasma with strong
vector interactions, $c_s = c$. Hence, the emission angle measurement
can yield information of the interactions in the plasma. This point
will be discussed in the following.
\end{enumerate}

\section{A hydrodynamical study of jet energy loss}

\begin{figure}[t]
\epsfig{file=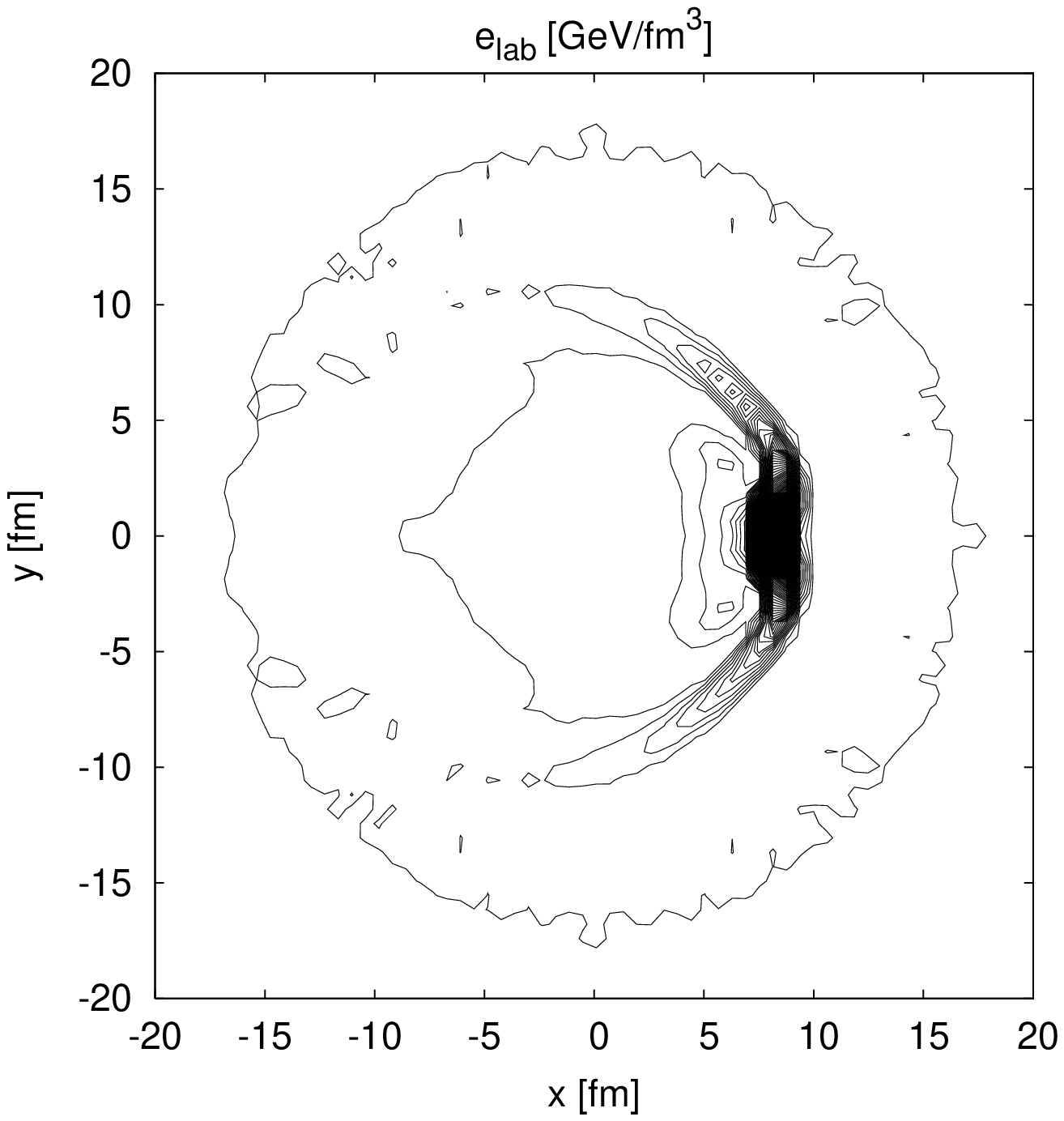,width=5.8cm}\hspace*{4mm}
\epsfig{file=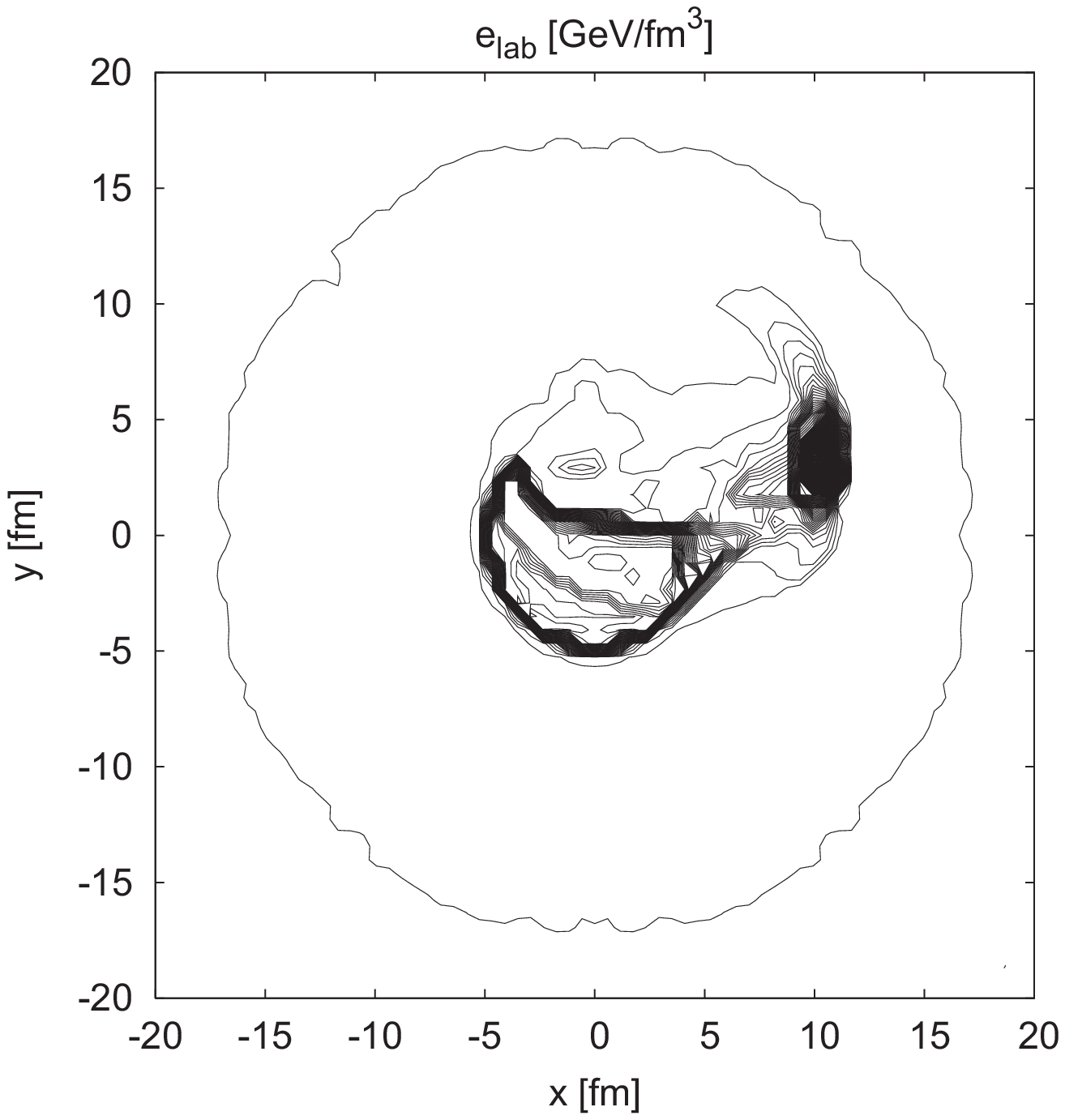,width=5.4cm}
\caption[]{Contour plot of the laboratory energy density at $t=12.8$
  fm/c for an ideal gas EoS (left) and for a hadron gas with
  first-order phase transition to QGP (right).}
\label{Jet21}
\end{figure}

% \begin{figure}[t]
% \begin{minipage}[t]{5.5cm}
% \epsfig{file=Contourlinien8.eps,width=0.8\textwidth}
% \caption[]{Contour plot of the laboratory energy density at $t=12.8$~fm/c
%   for an ideal gas EoS.}
% \label{Jet58}
% \end{minipage}
% \hspace*{1cm}
% \begin{minipage}[t]{5.5cm}
% \epsfig{file=Contourlinien02.eps,width=0.76\textwidth}
% \caption[]{Contour plot of the laboratory energy density at $t=12.8$~fm/c
%   for an hadron gas with first-order phase transition to QGP.}
% \label{Jet21}
% \end{minipage}
% \end{figure}

As mentioned in Sec. \ref{sectionjets}, the STAR and PHENIX
collaborations published the
observation\cite{StarAngCorr,Phenix,STAR} that the away--side jet in Au+Au
collisions for high-$p_T$ particles ($4 < p_T$(trigger)$< 6$~GeV/c,
$p_T$(assoc)$> 2$~GeV/c) with pseudo-rapidity $|y| < $~0.7 is
suppressed as compared to the away--side jet in p+p collisions (see
Fig. \ref{filimonov}).

This is commonly interpreted as parton energy loss, the so-called jet
quenching\cite{Sto04,jetquen}. One part of the back-to-back jet
created in the collision escapes (near-side jet), the other one
(away--side jet) deposits a large fraction of its energy into the dense
matter.

We use (3+1)dimensional ideal hydrodynamics, 
employ the (3+1)dimensional SHASTA (SHarp And Smooth
Transport Algorithm)\cite{SHASTA}, and follow the time evolution of a
fake jet that deposits its energy and momentum completely during a
very short time in a 2 fm$^3$ spatial volume of a spherically
symmetric expanding system.

The medium has an initial radius of $5$ fm, an initial energy density
of $e_0=1.68\, {\rm GeV/fm}^3$ and an initial profile velocity
increasing by radius as ${\bm v}(r) =0.02\,r/R$.

The initial energy density of the jet is increased by $\Delta e=5\,
\rm{GeV/fm}^3$ as compared to the medium and the jet 
material has an initial velocity of $v_x=0.96$~c.
We display the contour plots of the jet evolution at late state $t=12.8$~fm/c
for two different cases: Figure \ref{Jet21} (l.h.s) shows the evolution for an 
ultrarelativistic ideal gas EoS. Here, the jet is initially located 
in the region between $-5 \,{\rm fm}<x<3\,{\rm  fm},\,|$y$|<0.5\,{\rm fm},\, 
|$z$|<0.5\,{\rm fm}$. Figure \ref{Jet21} (r.h.s) depicts the evolution for a 
hadron gas with a first-order phase transition to QGP and a jet that
is initially located in the retion between $-3 \,{\rm fm}<x<-1\,{\rm
  fm},\,2.5 \,{\rm fm}<y<3.5\,{\rm  fm},|$z$|<0.5\,{\rm fm}$.

The jet-induced shock front and a deflection of the jet in case of an EoS 
with phase transition to QGP is clearly visible.

%------------------------------------------------------------
\section{Summary}

The NA49 collaboration has observed the collapse of both, $v_1$- and
$v_2$-collective flow of protons, in Pb+Pb collisions at 40 A GeV,
which presents evidence for a first-order phase transition in
baryon-rich dense matter. It will be possible to study the nature of
this transition and the properties of the expected chirally restored
and deconfined phase both at the HiMu/low energy and at the forward
fragmentation region at RHIC, with upgraded and/or second generation
detectors, and at the future GSI facility FAIR. According to lattice
QCD results\cite{Fodor04,Karsch04}, the first-order phase transition
occurs for chemical potentials above 400 MeV. Ref.\cite{Paech03} shows
that the elliptic flow clearly distinguishes between a first-order
phase transition and a crossover. Thus, the observed collapse of flow,
as predicted in Ref.\cite{Hofmann74,Hofmann76}, is a clear signal for
a first-order phase transition at the highest baryon densities.

A critical discussion of the use of collective flow as a barometer for
the EoS of hot dense matter at RHIC showed that
hadronic rescattering models can explain $< 30 \%$ of the observed
flow, $v_2$, for $p_T > 2$ GeV/c. We interpret this as evidence for
the production of superdense matter at RHIC with initial pressure way
above hadronic pressure, $p > 1$~GeV/fm$^3$.

The fluctuations in the flow, $v_1$ and $v_2$, should be
measured. Ideal hydrodynamics predicts that they are larger than 50 \%
due to initial state fluctuations. The QGP coefficient of viscosity
may be determined experimentally from the fluctuations observed.

We propose upgrades and second-generation experiments at RHIC, which
inspect the first-order phase transition in the fragmentation region,
i.e., at $\mu_B\approx~400$~MeV ($\sqrt{s}=4-12$ A GeV or $y \approx
4-5$ at full energy), where the collapse
of the proton flow analogous to the 40 A GeV data should be seen.

The study of jet-wake-riding potentials and bow shocks caused by jets
in the QGP formed at RHIC can give further clues on the EoS and
transport coefficients of the QGP.

\section*{Acknowledgements}
We like to thank M. Bleicher, I. Mishustin, K. Paech, H. Petersen,
D. Rischke and L. Satarov for stimulating discussions.

%-----------------------------------------------------------------


\begin{thebibliography}{120}


\bibitem{Fodor04}
    Z. Fodor and S. D. Katz,  JHEP {\bf 0203} (2002) 014;
    JHEP {\bf 0404} (2004) 050.
    
\bibitem{Karsch04}
    F. Karsch, J.\ Phys.\ G {\bf 30} (2004) S887;
    F.~Karsch,hep-ph/0701210.
    
\bibitem{Anishetty80}
     R.~Anishetty, Peter Koehler, and Larry~D. McLerran,
     Phys. Rev. {\bf D 22} (1980) 2793.
     
\bibitem{Date85}
     S.~Date, M.~Gyulassy, and H.~Sumiyoshi,
     Phys. Rev. {\bf D 32} (1985) 619.

\bibitem{Greiner:1987tg}
  C.~Greiner, P.~Koch, and H.~St\"ocker,
  Phys.\ Rev.\ Lett.\  {\bf 58} (1987) 1825;
  C.~Greiner, D.~H.~Rischke, H.~St\"ocker, and P.~Koch,
  Phys.\ Rev.\  D {\bf 38} (1988) 2797.   

 \bibitem{Koch86}
     P.~Koch, B.~M\"uller, and J.~Rafelski,
     Phys. Rept. {\bf 142} (1986) 167;
     I.~Zakout, C.~Greiner, and J.~Schaffner-Bielich,
     Nucl.\ Phys.\  A {\bf 781} (2007) 150;
     J.~Schaffner, C.~B.~Dover, A.~Gal, C.~Greiner, D.~J.~Millener, and
     H.~St\"ocker,
     Annals Phys.\  {\bf 235} (1994) 35;
     J.~Schaffner, C.~B.~Dover, A.~Gal, C.~Greiner, and H.~St\"ocker,
     Phys.\ Rev.\ Lett.\  {\bf 71} (1993) 1328.

\bibitem{Bratkov04}
    E.~L. Bratkovskaya {\it et al.},
    Phys.\ Rev.\  C {\bf 69} (2004) 054907.

\bibitem{Cleymans}
    J.~Cleymans and K.~Redlich,
     Phys. Rev. {\bf C60} (1999) 054908.

\bibitem{Bravina}
    L. V. Bravina {\it et al.},
    Phys. Rev.  {\bf C 60} (1999) 024904;
      Nucl. Phys.  {\bf A 698} (2002) 383.

\bibitem{BRAHMS_PRL03}
    I.~G. Bearden {\it et al.},
    Phys. Rev. Lett. {\bf 90} (2003) 102301.

\bibitem{Hofmann74}
    J.~Hofmann, H.~St\"ocker, W.~Scheid, and W.~Greiner,
  {Report of the Int. Workshop on BeV/Nucleon Collisions of Heavy Ions:
  How and Why}, Bear Mountain, New York, Nov. 29 - Dec. 1, 1974
  (BNL-AUI  1975).

\bibitem{Hofmann76}
    J.~Hofmann, H.~St\"ocker, U.~W. Heinz, W.~Scheid, and W.~Greiner,
    Phys. Rev. Lett. {\bf 36} (1976) 88.

\bibitem{Lacey}
  R.~A.~Lacey and A.~Taranenko, nucl-ex/0610029.

\bibitem{Landau}
    L.D. Landau and E.M. Lifshitz,
    \textit{Fluid Mechanics},
    Pergamon Press, New York, 1959.

\bibitem{Stocker79}
    H.~St\"ocker, J.~Hofmann, J.~A. Maruhn, and W.~Greiner,
    Prog. Part. Nucl. Phys. {\bf 4} (1980) 133.

\bibitem{Stocker80}
    H.~St\"ocker, J.~A. Maruhn, and W.~Greiner,
    Phys. Rev. Lett. {\bf 44} (1980) 725.

\bibitem{Stocker81}
  H.~St\"ocker {\it et al.},
  Phys. Rev. Lett. {\bf 47} (1981) 1807.

\bibitem{Stocker82}
    H.~St\"ocker {\it et al.},
    Phys. Rev. {\bf C 25} (1982) 1873.

\bibitem{Stocker86}
    H.~St{\"o}cker and W.~Greiner.
    Phys. Rept.  {\bf 137} (1986) 277.

\bibitem{Csernai99}
    L.~P. Csernai and D.~R\"ohrich,
    Phys. Lett. {\bf B 458} (1999) 454.

\bibitem{Csernai04}
    L.~P. Csernai {\it et al.},
    hep-ph/0401005.

\bibitem{Rischke:1995pe}
  D.~H.~Rischke, Y.~P\"urs\"un, J.~A.~Maruhn, H.~St\"ocker, and W.~Greiner,
  Heavy Ion Phys.\  {\bf 1} (1995) 309.

\bibitem{Schmidt93}
    W.~Schmidt {\it et al.},
     Phys. Rev. {\bf C 47} (1993) 2782.

\bibitem{Muronga01}
    A. Muronga,
    Heavy Ion Phys. {\bf 15} (2002) 337.

\bibitem{Muronga03}
    A. Muronga,
    Phys. Rev. {\bf C 69} (2004) 034903.
    
\bibitem{Romatschke}    
    R.~Baier and P.~Romatschke, arXiv:nucl-th/0610108;
    R.~Baier, P.~Romatschke and U.~A.~Wiedemann, Phys.\ Rev.\  C {\bf 73}, 064903 (2006);
    R.~Baier, P.~Romatschke and U.~A.~Wiedemann, Nucl.\ Phys.\  A {\bf 782}, 313 (2007);
    P.~Romatschke, arXiv:nucl-th/0701032;
    P.~Romatschke and U.~Romatschke, arXiv:0706.1522 [nucl-th].
    
\bibitem{Mota}
   T.~Koide, arXiv:nucl-th/0703038;
   T.~Koide, G.~S.~Denicol, Ph.~Mota and T.~Kodama, Phys.\ Rev.\  C {\bf 75} (2007) 034909;
   Ph.~Mota, G.~S.~Denicol, T.~Koide and T.~Kodama, arXiv:hep-ph/0701162.
   
\bibitem{HeinzChaudhuri}
    U.~W.~Heinz, arXiv:nucl-th/0512051.
    A.~K.~Chaudhuri and U.~W.~Heinz, J.\ Phys.\ Conf.\ Ser.\  {\bf 50}, 251 (2006); 
    U.~W.~Heinz, H.~Song and A.~K.~Chaudhuri, Phys.\ Rev.\  C {\bf 73}, 034904 (2006);
    A.~K.~Chaudhuri, arXiv:nucl-th/0703027;
    A.~K.~Chaudhuri, arXiv:nucl-th/0703029.

\bibitem{Hirano}
 T.~Hirano, U.~W.~Heinz, D.~Kharzeev, R.~Lacey and Y.~Nara, Phys.\ Lett.\  B {\bf 636}, 299 (2006).
 
\bibitem{Brachmann97}
    J.~Brachmann {\it et al.},
    Nucl. Phys. {\bf A 619} (1997) 391.

\bibitem{Hofmann99}
    M.~Hofmann {\it et al.},
    nucl-th/9908031.

\bibitem{Hartnack89}
    C.~Hartnack {\it et al.},
    Nucl. Phys. {\bf A 495} (1989) 303c.

\bibitem{Bass98}
    S.~A. Bass, M.~Gyulassy, H.~St{\"o}cker, and W.~Greiner,
    J. Phys. {\bf G 25} (1999) R1.

\bibitem{Cassing99}
    W.~Cassing and E.~L. Bratkovskaya,
    Phys. Rept. {\bf 308} (1999) 65.

\bibitem{Weber02}
    H.~Weber, E.~L. Bratkovskaya, W.~Cassing, and H.~St\"ocker,
    Phys. Rev. {\bf C 67} (2003) 014904.

\bibitem{Andronic03}
    A.~Andronic {\it et al.},
    Phys. Rev. {\bf C 67} (2003) 034907.

\bibitem{Andronic01}
    A.~Andronic {\it et al.},
    Phys. Rev. {\bf C 64} (2001) 041604.

\bibitem{Soff99}
    S. Soff, S. A. Bass, M. Bleicher, H. St\"ocker, and W. Greiner,
    nucl-th/9903061.
    
\bibitem{Sahu1}
    P. K. Sahu and W. Cassing,
     Nucl. Phys. {\bf A 672} (2000) 376.

\bibitem{Sahu2}
    P.~K.~Sahu and W.~Cassing,
     Nucl. Phys. {\bf A 712} (2002) 357.
    
\bibitem{Sto04}
	H.~St\"ocker,
	Nucl. Phys. \textbf{A 750} (2005) 121. 

\bibitem{Brach00}
    J.~Brachmann,
    PhD thesis, J. W. Goethe - Universit\"at Frankfurt am Main, 2000.

\bibitem{Brach99}
    J.~Brachmann {\it et~al.},
     Phys. Rev. {\bf C 61} (2000) 024909.
     
\bibitem{Paech00}
    K.~Paech, M.~Reiter, A.~Dumitru, H.~St\"ocker, and W.~Greiner,
     Nucl. Phys. {\bf A 681} (2001) 41.

\bibitem{NA49_v2pr40}
    C.~Alt {\it et al.},
    Phys. Rev. {\bf C 68} (2003) 034903.

\bibitem{Petersen}
    H.~Petersen, Q.~Li, X.~Zhu and M.~Bleicher,
  Phys.\ Rev.\  C {\bf 74}, (2006) 064908.
  
\bibitem{Andronic00}
    A.~Andronic {\it et al.},
     Nucl. Phys. {\bf A 679} (2001) 765.

\bibitem{Andronic99}
    A.~Andronic,
    Nucl. Phys. {\bf A 661} (1999) 333.

\bibitem{Paech03}
    K.~Paech, H.~St\"ocker, and A.~Dumitru,
    Phys. Rev. {\bf C 68} (2003) 044907;       
    Phys. Rev. {\bf C 62} (2000) 064611.

\bibitem{StarAngCorr}
    C.~Adler {\it et~al.} [STAR collaboration],
     Phys. Rev. Lett. {\bf 90} (2003) 082302;
     L.~Molnar, nucl-ex/0701061.

\bibitem{Adl03b}
	C.~Adler {\it et~al.} [STAR collaboration],
	Phys. Rev. Lett. \textbf{91} (2003) 072304;
	S.~S.~Adler {\it et al.}  [PHENIX collaboration], 
	Phys.\ Rev.\  D {\bf 74} (2006) 072002.

\bibitem{Wan04}
	Fuqiang Wang [STAR collaboration],
	J.\ Phys.\ G {\bf 30} (2004) S1299;
	J.~G.~Ulery and F.~Wang, nucl-ex/0609017;
	F.~Wang, Nucl.\ Phys.\  A {\bf 783} (2007) 157;
	F.~Wang, nucl-ex/0610027.

\bibitem{Jac05}
	B. Jacak [PHENIX collaboration], N.N. Ajitanand [PHENIX collaboration],
	talks at Int. Conf. on Physics and Astrophysics
	of Quark Gluon Plasma, Kolkata, India, 2005;
	B.~Jacak  [PHENIX collaboration], J.\ Phys.\ Conf.\ Ser.\
        {\bf 50} (2006) 22.


\bibitem{Rischke90}
    D.~H. Rischke, H.~St\"ocker, and W.~Greiner,
    Phys. Rev. {\bf D 42} (1990) 2283.

\bibitem{Cha86}
	G.F. Chapline and A. Granik,
	Nucl. Phys. \textbf{A 459} (1986) 681.

\bibitem{Gla59}
	A.E. Glassgold, W. Heckrotte, and K.M. Watson,
	Ann. Phys. \textbf{6} (1959) 1.

\bibitem{Kho80}
	V.A. Khodel, N.N. Kurilkin, and I.N. Mishustin,
	Phys. Lett. \textbf{B 90} (1980) 37.

\bibitem{Cas04}
	J. Casalderrey--Solana, E.V. Shuryak, and D. Teaney,
	J.\ Phys.\ Conf.\ Ser.\  {\bf 27} (2005) 22;
	J.~Casalderrey-Solana, hep--ph/0701257;
	F.~Antinori and E.~V.~Shuryak, J.\ Phys.\ G {\bf 31} (2005) L19.

\bibitem{Ruppert}
        T.~Renk and J.~Ruppert, Phys.\ Lett.\  B {\bf 646} (2007) 19;
        T.~Renk and J.~Ruppert, arXiv:hep-ph/0701154;
        T.~Renk and J.~Ruppert, arXiv:hep-ph/0702102.
	
\bibitem{Chaudhuri}
        A.~K.~Chaudhuri and U.~Heinz, Phys.\ Rev.\ Lett.\  {\bf 97}, 062301 (2006).

\bibitem{Satarov}
  L.~M.~Satarov, H.~St\"ocker, and I.~N.~Mishustin
  Phys.\ Lett.\  B {\bf 627} (2005) 64.

\bibitem{Gal04}
	W. Cassing, K. Gallmeister, and C. Greiner,
	J. Phys. \textbf{G 30} (2004) 801.


\bibitem{CGG}
    W.~Cassing, K.~Gallmeister, and C.~Greiner,
    { Nucl. Phys.} {\bf A 735} (2004) 277.

\bibitem{Phenix}
  A.~Sickles  [PHENIX collaboration], nucl-ex/0702007.


\bibitem{Baum75}
    H.~G. Baumgardt {\it et~al.},
    Z. Phys. {\bf A 273} (1975) 359.	
    
 \bibitem{Ruppert:2005uz}
   J.~Ruppert and B.~M\"uller,
   Phys.\ Lett.\  B {\bf 618} (2005) 123.


\bibitem{Kharzeev}
    D.~Kharzeev,
    private communication.

\bibitem{Schafer78}
    W.~Sch\"afer, H.~St\"ocker, B.~M\"uller, and W.~Greiner,
    Z. Phys. {\bf A 288} (1978) 349.

\bibitem{STAR}
    J.~Adams {\it et al.}  [STAR collaboration], Phys.\ Rev.\ Lett.\
    {\bf 91} (2003) 072304.
    
\bibitem{jetquen}
    M.~Gyulassy, P.~Levai, and I.~Vitev, Nucl.\ Phys.\ B {\bf 594}
    (2001) 371.

\bibitem{SHASTA}
    D.~H.~Rischke, S.~Bernard, and J.~A.~Maruhn, 
    Nucl.\ Phys.\ A {\bf 595} (1995) 346;
    D.~H.~Rischke, Y.~P\"urs\"un, J.~A.~Maruhn, H.~St\"ocker and W.~Greiner,
    Heavy Ion Phys.\  {\bf 1}, 309 (1995).

        
\end{thebibliography}
\end{document}